# Augmentation Technologies and AI—An Ethical Design Futures Framework

By Ann Hill Duin and Isabel Pedersen

## *Overview*

Augmentation technologies, fueled by artificial intelligence (AI), are undergoing a process of adaptation and normalization geared to everyday users in various roles as practitioners, educators, and students. While new innovations, applications, and algorithms are developed as "augmentation technology," Chapter 1 focuses on human subjects, contexts, and rhetorical strategies proposed for them by external actors. The chapter discusses core functions of technical and professional communication and provides rationale for positioning technical and professional communicators (TPCs) to understand augmentation technologies and AI as a means to design ethical futures across this work. An overview of *Augmentation Technologies and AI—An Ethical Design Futures Framework* serves as a guide for reframing professional practice and pedagogy to promote digital and AI literacy surrounding the ethical design, adoption, and adaptation of augmentation technologies. The chapter concludes with an overview of the remaining chapters in this book.

## *Key Questions*

- How might augmentation technologies best be defined?

● What powerful rhetorics signal the augmentation technology and AI promise of augmented humans, cyborgs, and posthumans?

● The progress of augmentation technologies and AI follows decades of theoretical research on cyborgs, posthumanism, transhumanism, and non-humanism in philosophy, critical theory, critical feminism, and feminist new materialism. How might technical and professional communicators use this research for building understanding of augmentation technologies and AI?

● How will core functions of technical and professional communication—understanding audience, user experience, content development, collaboration, and design—transform alongside augmentation technologies and AI?

## *Chapter 1 Links*

Throughout the chapter, we refer to articles, videos, and reports which can be found in a related chapter collection at Fabric of Digital Life called *Augmentation Technologies and AI—An Ethical Design Futures Framework* (Duin & Pedersen, 2022). You can find a link to this collection at https://fabricofdigitallife.com.

## *Introduction*

The Press Release for Gartner Inc.'s Hype Cycle for Emerging Technologies (2020) describes 30 "must-watch" technologies. Amid descriptions for changing virtual workplaces and solutions for the global pandemic crisis appears a seemingly mundane reference to technologies that will "alter the state of your brain." The release explains, "the way people interact with the digital world is also moving beyond screens and keyboards to use a combination of interaction modalities (e.g. voice, vision, gesture), and even directly altering our brains." Chris Duffey (2019), writing in preparation for the CES 2020 Las Vegas exhibition, asks, "What if we could tap into superhuman powers to be better at school, excel at sports, succeed in business and ultimately live a longer and fuller life? Who wouldn't want that competitive edge?" And corporate innovation announcements reflect this trend: "Elon Musk says Neuralink could start planting computer chips in human brains within the year," emphasizing the trials that will be conducted to shift brain implants from a speculative proposed technology to one ready for adoption (Kay, 2021).

The purpose of this book is to cultivate an even deeper understanding of human augmentation and AI technology, build technical and professional communication capacity to articulate its benefits and risks, and provide direction for future practice and collaboration. This book is concerned with augmentation technologies and AI that will alter the brain, but it is also concerned with recent discourses that are so ready to promote brain augmentation as a paradigm shift. Katina Michael and colleagues (2020), in their review of the rise of implantables, emphasize that "in some ways, the end user is the new 'last mile' in the global interconnected network topology" (p. 97). With this point, we argue that augmentation technologies fueled by AI are undergoing a process of normalization geared to everyday users in various roles as practitioners, caregivers, educators, and students.

For example, regarding practitioners, a 2021 Massachusetts Institute of Technology (MIT) Sloan Management Review (SMR) survey, "The Workforce Ecosystem Perspective," suggests that the meaning of the term "workforce" is undergoing a rapid shift, due in part to emerging augmentation technologies. Of the 5,118 global executives surveyed, MIT SMR found that 87% of respondents "consider their workforce to encompass more than their employees"; 37% view "technology for workforce augmentation" as "part of [their organization's] workforce"; and 56% indicate that this category will increase in the next two years. However, despite the trend toward increasing adoption of augmentation technologies and AI in the workplace, only a minority of these global executives believe their organization is adequately prepared to manage a future workforce made up of so many "external participants."

Given this workforce context, MIT SMR (2022) proposes that executives replace the traditional employee framework with "a workforce ecosystem approach." This would require new management strategies and a reconsideration of underlying philosophies, systems, and processes. They provide specific guidance for workforce planning, talent acquisition, performance management, compensation and rewards, learning and development, career paths, and organizational design. Similar to how MIT SMR addresses the need to adapt management approaches to the changes that augmentation technologies bring to organizations, technical and professional communication practitioners, instructors, and researchers will need to adapt to similar changes currently being normalized.

Part of this process of normalization includes narratives that explore ambiguities tightly bound with the pursuit of augmentation. *Black Mirror*, created by Charlie Brooker (2011–2019), is a popular science fiction television series that deals with augmentation in stark dystopian terms. Reminiscent of the near future worlds of *Ex Machina* and *Her*, *Black Mirror* often concentrates on the harsh consequences of implanting humans with computer chips. In season one, episode 3, "The Entire History of You," characters have a "grain" implanted behind their ears that records everything they see and hear. Using a remote, they can play back memories on screens. It illustrates an enduring real-life ambition for technologies to augment human memories by preserving all in digital archives. In another episode, a doctor receives a neurological implant that allows him to feel the physical pain of his patients, thus being able to diagnose them more accurately through an augmented sense of empathy.

*Black Mirror* stands in contrast to the celebratory futurism of Marvel's cinematic universe franchise of superhero films or even the uber-cool urbanism of the *Matrix* trilogy (Wachowski &

Wachowski, 1999–2003). It critiques tactics of current technology adoption based on providing work efficiencies that replace human work or activities with autonomous agents. It sometimes sets the narratives in raw natural landscapes, emphasizing the non-digital materiality of life and uncontrollable environments that resist human technological interference. *Black Mirror* emphasizes socio-political contexts that position augmentation technologies as instigators of inequalities in society—the digital divide. Relevant as well is the way it frames its narratives on current innovations in research and development. Empathetic AI and augmenting human empathy are two real-world technologies under development (Dial, 2018; Wu, 2019): Rachel by Emoshape uses AI algorithms to respond to meaning and language around her in real time, then expresses herself accordingly (Emoshape Inc., 2018). *Black Mirror* not only critiques current emerging human augmentation technologies, it also scrutinizes Big Tech's interest in this sector as well as mainstream media's recent preoccupation with reporting on it.

The discourses that justify these technologies use powerful rhetorics promising to deliver transparency in AI decision-making along with augmented humans, cyborgs, and posthumans as a tactic for adoption. We also acknowledge that many individuals self-identify as augmented or aspire to augment themselves using technologies, including those who have garnered some kind of fame (each of the following is included in the Fabric of Digital Life collection, *Augmentation Technologies and AI—An Ethical Design Futures Framework*): Stelarc (n.d.); Rob Spence (Spence & Jaworski, n.d.); and cyborg activists Neil Harbisson (n.d.), Kevin Warwick (infonomia, 2008), and Moon Ribas (n.d.). Their autobiographical narratives weave throughout discourses about technologies. Biohacking, the lifestyle practice of using science and technology to improve one's physicality or cognitive abilities, thus becomes more mainstream. Biohackers are discussed in popular science venues, popularized by Silicon Valley titans. One *Vox* article describes it as "one branch of transhumanism, a movement that holds that human beings can and should use technology to augment and evolve our species" (Samuel, 2019).

### *Augmented Humans: Histories, Theories, and Rhetorics*

The progress of augmentation technologies follows decades of relevant theoretical research on cyborgs, posthumanism, transhumanism, and non-humanism in philosophy, critical theory, critical feminism, and feminist new materialism (Haraway, 1985; Hayles, 1999; Barad, 2003; Wolfe, 2010; Braidotti, 2013). Transhumanism challenges bioethical positions due to its encouragement of human enhancement as a concept, a moral stance that continues to reverberate across augmentation discourse.

Theories of cognitive augmentation are often rooted in cybernetics theory. Influenced by Maurice Merleau-Ponty, Andy Clark proposed the "extended mind" theory in 1997, leading to a tradition that is in many ways key to augmentation technology strategies today. Clark writes of evolution as:

> a process that must build new solutions and adaptive strategies on the basis of existing hardware and cognitive resources. And it is empowered, as we have seen, by the availability of a real-world arena that allows us to exploit other agents, to actively seek useful inputs, to transform our computational tasks, and to offload acquired knowledge into the world.
>
> (1997, p. 88)

In later work he offers a nuanced take on extended mind to predict progress toward cyborgism:

> We shall be cyborgs not in the merely superficial sense of combining flesh and wires but in the more profound sense of being human-technology symbionts: thinking and reasoning systems whose minds and selves are spread across biological brain and nonbiological circuitry.
>
> (Clark, 2003, p. 3)

N. Katherine Hayles (2002), prominent posthuman theorist, comments on Clark's reasoning and states, "Agency still exists, but for the posthuman it becomes a distributed function" (p. 319). She clarifies the subject's lived experience: "Living in a technologically engineered and information-rich environment brings with it associated shifts in habits, postures, enactments, perceptions—in short, changes in the experiences that constitute the dynamic lifeworld we inhabit as embodied creatures" (p. 299). Throughout this book, we draw on these points of view to explain augmentation technologies across various forms of enhancement, as well as directions to frame ways to presuppose their emergence across the platforms now under development and being used.

Historically, enthusiast groups and organizations connected to human enhancement have driven popular and oftentimes controversial momentum for augmentation as a positive, ambitious goal for humanity. These movements contribute terminology and motivate people to consider issues that might threaten people's value systems or even legal systems. Radical life extension, explored by Ray Kurzweil and Terry Grossman (2009), has pushed ethical boundaries for some, impacting the uptake of certain technologies. Groups like Humanity+ (also known as the World Transhumanist Association), Extropy Institute, and Singularity University feature speakers, scientists, researchers, lawmakers, and thought leaders, according to each group's agenda. This book does not focus on or promote these groups or these movements; however, we acknowledge them as motivators for addressing human augmentation in public and professional spheres.

Integrated with philosophical and existential categories for human-machine mergers important to the idea of *augmenting* humans are rhetorical interventions that cause people to adopt or reject them. Also important to this book is Kevin Thayer's (2014) work "Mapping Human Enhancement Rhetoric," in which he ponders how "questions can be raised about the shifting ethical positions in human and transhuman enhancement discourse" (p. 31). Gregory Hansell and William Grassie (2011) further identify transhumanism as a value system or belief structure with an ideological foundation trusting and expecting that dramatic human enhancement is possible:

The applied sciences involved include dramatic advancements in the neurosciences, genomics, robotics, nanotechnology, computers and artificial intelligence. In some combination of the above bioengineering, transhumanists imagine the possibilities in the near future of dramatically enhancing human mental and physical capacities, slowing and reversing the aging process and controlling our emotional and mental states. The imagined future is a new age in which people will be freed from mental disease and physical decrepitude, able to consciously choose their 'natures' and those of their children. At first glimpse, it all seems like a wonderful thing, life lived more abundantly, but Francis Fukuyama calls this transhumanist vision 'the most dangerous idea in the world.' (p. 13).

One distinguisher of transhumanism is the value of human control over emotional and mental states. Human control is a contested value of current and proposed augmentation and one that we explore throughout this book in light of automation.

Following several of these premises, Isabel Pedersen and Tanner Mirrlees (2017) argue that "transhumanism reveals four predominant and often sensationalized themes used to promote human technological advancement as a seemingly obvious future in the mainstream" (p. 40). They argue that transhumanist rhetoric is usually buttressed with (1) a legitimate claim of technological or scientific progress; (2) a claim based on human agency or human control; (3) a claim for superheroism or a superhuman ability; and (4) an urgent claim based on a need or issue over human vulnerability. Using the example of exoskeleton development for physical, non-medical human enhancement, they conclude that "exciting news about technological innovations thrives across networked channels. People consume and circulate popular techno news as entertainment to tantalize, to scandalize, and to authorize, but as a result, society simultaneously accepts values in terms of this hyped technology" (p. 44).

Such public sensationalism along with fear become governing rhetorics of augmentation technologies. The development of a so-called superintelligence has fueled mainstream angst over the danger of evolving an Artificial General Intelligence (AGI) for humans. This kind of public concern led Silicon Valley entrepreneur Elon Musk to perpetuate the myth that humans ought to enhance themselves to keep up with AI. An element of this book is the monitoring of such public justifications for technology under extremely hyped or even illogical conditions. In many ways the rhetoric of augmentation technology and AI is also technoliberal rhetoric:

> Technoliberal rhetoric is most visible as it is articulated to digital technologies, diffusing the terms, tropes, and frames of technoliberalism through sites of tech evangelism like Apple Keynotes, TED Talks, CES (the technology trade show formerly known as the Consumer Electronic Show) and through commercial advertisements across different media.

> (Pfister & Yang, 2018, p. 241)

We identify technoliberalism as "the condition of digitality [that] intensifies neoliberal governing rationalities in the context of public sphere deliberation…. Technoliberalism may be the default public philosophy of digital culture, but it is neither natural nor inevitable" (Pfister & Yang, 2018, p. 250).

Writing about communication, machines, and human augmentics, John Novak and colleagues (2016) ask, "When do electronic tools cease to be 'simply' tools, and become meaningfully part of ourselves?" From this pivotal question, they add "When might we think of these tools as augmenting ourselves, rather than simply amplifying our capabilities?" Echoing Mark Weiser's 1991 claim of ubiquitous computing, that society's inclination is to "push computers into the background," augmentics is contextualized in ontological terms. Robert Kenyon and Jason Leigh (2011) adopted the term "human augmentics" to describe "a call to arms for the rehabilitation [medical] community to think outside of their boundaries—to think of the problem in terms of a larger interconnected ecosystem of augmented humans rather than a patchwork of disconnected sub-systems" (p. 6761). Zizi Papacharissi (2019) uses the term "augmentics" to describe a contested state relating to expanded human capability: "By focusing on the theme of augmentics, I aimed at directing our attention to ways in which technologies expand our capabilities" (p. 7). Rejecting a state of assumed betterment, Papacharissi settles on difference, stating that "we will not become better, superior, or more advanced—that will be a function of how we put technology to use, and ultimately, a call that will be forever subjective. We will, or rather, we have the opportunity to, become different" (p. 7). Useful too in Papacharissi's collection is Douglas Guilbeault and Joel Finkelstein's (2019) interpretation of augmentics. Following Marshall McLuhan, they envision it within the cybernetic turn and use ecological principles to explore selves in decentralized contexts, articulating the possibility for an "augmented collective consciousness" (p. 179).

Leading rhetorical and professional communication scholars Heidi McKee and James Porter, in their 2017 chapter "AI Agents as Professional Communicators," acknowledge that "increasingly, humans are communicating with AI agents, often without knowing they are doing so. The implications of AI for professional communication and for organizations and business professionals who deploy AI agents are profound" (p. 135). In their exploration of rhetorical and ethical issues in AI within professional communication, their overarching question is "can AI bots be effective professional communicators?" (p. 135). Based on their case studies of use of AI agents, they articulate the most "interesting complexity of AI for rhetorical interaction" as occurring "when AI agents aim to become more 'intelligent' and when interactions become potentially even more open-ended, existing in the rhetorical free-for-all that is, ultimately, human conversation" (p. 153). Chapter 7 of this volume includes examples of digital employees designed for such communication.

Continuing with focus on rhetorical context for AI writing, in their 2020 work McKee and Porter emphasize that AI writing systems are "built on an information transfer model of communication that assumes text production is a simple matter of converting raw data into sentences and paragraphs" (p. 110), and that "When humans and AI systems interact, miscommunication occurs and ethical issues arise from lack of understanding about [social and rhetorical] context" (p. 113). They offer two ethical principles to guide design of AI writing systems, principles that we align with throughout this book:

- An ethic of transparency: humans must know the rhetorical context and whether they are interacting with an AI agent—whether in mobile text, social media, or other communication (p. 113); and

- An ethic of critical data awareness: a methodological reflexivity about rhetorical context and omissions in the data that need to be provided by a human agent or accounted for in machine learning (p. 110).

And in their 2022 study of "Team Roles & Rhetorical Intelligence in Human-Machine Writing," McKee and Porter continue to acknowledge the "immense effect" that fast-evolving AI writing technologies have on professional communication. While most current AI writing systems are "resource tools or assistants," they note that in "bounded verticals" AI writing agents function as "higher-contributing team members . . . who are . . . acting as *writers*" (p. 3). They again note the critical importance of understanding the rhetorical context for communication, with a call for considering "rhetorical intelligence" for AI. Moreover, they stress that "we are on the onset of a seismic change in writing and teaming for professional communication" (p. 390), with which we strongly agree. Critical to navigating this change is understanding of augmentation technologies, as both a rhetorical and ethical phenomenon.

## *Augmentation Technology Definition*

Isabel Pedersen and Andrew Iliadis (2020) define "embodied computing" as those technologies that "exist in topographical [on the body], visceral [in the body], and ambient [around the body] relationships with the body" (p. xi). As a kind of embodied computing technology, augmentation technologies and AI enhance human capabilities or productivity by adding to the body (or ambient environment around the body) cognitive, physical, sensory, and/or emotional enhancements. Examples include brain–computer interaction devices for cognitive enhancement; robots marketed to improve human social interaction; wearables that extend human senses, augment creative abilities, or overcome physical limitations; implantables that amplify intelligence or memory; devices or algorithms for affective computing; Internet of Bodies (IoB) and Internet of Things (IoT) for ambient interaction or surveillance with places/spaces; and extended reality (XR) technologies, e.g., augmented reality (AR) and virtual reality (VR), to alter human interaction with people's lived reality. Judith Hurwitz and colleagues, longtime scholars of AI, write, "The most pragmatic and useful way to benefit from AI and machine learning is to implement these powerful technologies as an augmentation to human intelligence" (Hurwitz et al., 2020, p. xix).

Most importantly, we redefine augmentation technologies and AI as social and rhetorical phenomena. For example, social robots and virtual assistants are slated to augment human experiences and relationships, to communicate with, care for, monitor emotions of, entertain, instruct, and supervise humans, and to assist in teaching and practice. Human experiences, social life, arts and human identities, and the practice of technical and professional communication are affected by this momentum (Hartzog, 2015; Slane et al., 2020;).

Augmentation technologies work to enhance the way we do things and perceive our environment. Therefore, human augmentation (Human 2.0), once assumed to focus solely on extending human capabilities, now works to transform human abilities for nonmedical reasons. According to an expert panel from the Forbes Technology Council (2020),

Human augmentation technology offers vast potential for many different industries. These developments combine medicine and technology to increase the capabilities of the human body. However, those outside the tech industry may not even be aware of how much augmentation can enhance our daily lives.

For example, Scribe.ai is working to "supplement our neurons to make even the most mundane occurrences into something unforgettable. . . . The plan is to methodically capture and store all sorts of data—audio, video, and eventually biometric—that can be easily searched or cleverly invoked in a way that augments your actual memory." Scribe.ai's first product is an add-on to Zoom in which "a faceless participant . . . a dynamic rapporteur" would join a meeting to log "what people say and what they look like as they say it," which then might be used to augment memory and future work in more immersive ways (see Figures 1 and 2). This work is described in the article "A New Company Pursues Total Recall—Starting With Zoom," by Steven Levy (2021).

Figure 1.1. Any participant may invite Scribe to join the meeting, start recording, transcribe what is said, summarize key points, and search all past transcripts for key points. Used with permission of Dan Siroker, CEO of Scribe.

Figure 1.2. Scribe's user dashboard. Used with permission of Dan Siroker, CEO of Scribe.

Augmentation technologies are alluring and celebrated. They are afforded the freedom to operate as if necessary and imminent rather than speculative because of their shocking facility to enhance human abilities and practices. At the same time, they evince a significant lack of forethought, governance, societal, or consumer control. Augmentation technologies emerge amid fluctuating corporate and international spheres that have not yet been regulated. For example, Moodmetric and Oura smart rings are two products designed with self-tracking capabilities and marketed with the promise of promoting wellbeing. In "Making sense with sensors: Self-tracking and the temporalities of wellbeing," Martin Berg (2017) analyzed rhetoric used to explain the functionality of these rings. In marketing materials and user manuals, Berg found language about the impenetrable nature of the body's natural signals, conceptions of the body as a machine that needs "optimizing," and assumptions about accelerated time in modern society.

Moodmetric and Oura are part of what researchers Mark Andrejevic and Mark Burdon (2014) call the sensor society. In a 2015 TED talk, Burdon discussed corporate rhetoric used to sell sensorized objects, but he also warned of ethical issues associated with these new products connected to large data collection systems. Now is the time, Burdon argues, to ask fundamental questions about how such devices can be designed and used ethically.

Consumers are already adopting the technology, and the industry is growing rapidly (Kundan, 2021). At CES 2022 in Las Vegas, many new digital health devices with sensor technology were on display. Abbott Laboratories, a medical device company, announced a push into the area of biowearables during the first CES keynote address featuring a healthcare company. In the Abbot presentation, CEO Robert Ford spoke about his vision for "human-powered health." Biowearables "will be like having a window into your body," Ford explained. This technology offers "science that you will be able to access any time so you can understand what your body is telling you and what it needs" (Abbott Health, 2022). Ford concluded the address with a vision for a future with sensors that will empower health consumers, give them more control, more freedom, and less disruption to daily life.

The idea that sensors can provide a window into the body is a common rhetorical phrase used to explain wearable sensor devices. Berg reported encountering this kind of language frequently in his research. Moodmetric and Oura are just two examples of self-tracking devices supposedly designed to promote wellbeing by monitoring body signals that were previously undetected while a person was unaware or sleeping. Though some researchers refute the claim that sleep tracking supports better sleep habits (Zraick et al., 2019), promoters of these devices point to the evidence of millions of satisfied customers. Promotional videos for Moodmetric and Oura support Berg's research concerning the corporate rhetoric that continues to be associated with such products. Further, people are socialized to want to "review data about themselves collected by other actors, such as social media metrics, employee dashboards, educational outcomes, medical records and so on" (Lupton, 2018, p. 1). They are willing to provide the labor and biometric data needed to fuel these platforms. Deborah Lupton (2016; 2018; 2020) has contributed much on the quantified self movement and the way devices "work to capture and materialise immanent dimensions of human embodiment, creating human–data assemblages" (Lupton, 2018, p. 1).

Corporate rhetoric drives adoption before people can understand its impact. As if they were already legitimate, AI-based technologies are driven by their promised transformative claims to disrupt traditional domains. Clearview AI provides an especially representative example of such a dramatic transformation. Originally a small AI startup, Clearview AI scraped three billion public images from Facebook's servers to provide law enforcement agencies with a significant algorithm to identify criminals (Hill, 2020). A vast international collection of people's social media images instantly became a pool of potential suspects. Clearview AI's business model not only exceeded privacy regulations and national laws, it challenged large social media corporations' abilities to control their own platforms. Facebook could no longer control how users would *be used by* their own technology.  Based on a lawsuit in Illinois, in May 2022, Clearview AI agreed in a settlement to stop selling its massive database of images. As discussed by Greg Bensinger (2022),

> The Biometric Information Privacy Act of Illinois sets strict limits on the collection and distribution of personal biometric data, like fingerprints and iris and face scans. The Illinois law is considered among the nation's strongest, because it limits how much data is collected, requires consumers' consent and empowers them to sue the companies directly, a right typically limited to the states themselves.

Therefore, as Clearview AI and other technology companies profit by deploying public images to law enforcement and other private entities, this lawsuit shows that "effective statutes can help bring some of Big Tech's more invasive practices to heel" (Bensinger, 2022).

This is but one example of how augmentation technologies and AI have been emerging over recent decades driven by corporate development, university research, military-industrial complex development, increased biometric data availability, new AI techniques, biological technologies, upgraded computing power, and maturing digital architectures. The emergence of AI has recently accelerated through machine-learning algorithms, natural language processing, and predictive models that inform the design of technologies. The *Next Generation of Emerging Global Challenges Horizons 2030* report emphasizes that "the increasing sophistication of physical and cognitive augmentation technologies will unlock new potential for human abilities, health and longevity, potentially raising divisive social, legal, and psychological issues" (Policy Horizons Canada, 2018). Driven largely by such corporate research advancement, technology development, increased data availability, the rise of AI, upgraded computing power, and new architectures, augmentation technologies are emerging largely unchecked and largely not understood by technical communication professionals and scholars/instructors. AI rhetoric contributes to the hype, and policy makers are taking note of it.

This book addresses the expectations emanating from these developments and the proliferation of large companies promising innovations as mainstream phenomena. Elon Musk argues that humans communicate and think too slowly as he markets Neuralink Corporation's development of brain-implantable neurolace (Ricker, 2016; Orth, 2020). One result is that this kind of corporate-driven efficiency worldview becomes instantiated within the public sphere discourse about technology. It neglects communities, misrepresents cultures, and even harms individuals, whether these technologies are adopted or not (Crawford et al., 2019).

We recognize that global actors constantly alter the way augmentation technologies and AI are adopted, impacting the future and the implications for technical and professional communicators. This book magnifies instances of augmentation technologies and AI in order to appropriately contextualize digital enhancements. We also acknowledge that "algorithmic systems can cause harm when they fail to work as specified—i.e., in error—but may just as well cause real harms when working *exactly* as specified" (Moss et al., 2021).

Moreover, industry reports such as Jim Guszcza, Harvey Lewis, and Peter Evans-Greenwood's (2017) *Deloitte Insights* argue that "humans and computers think better together." In *Writing Futures: Collaborative, Algorithmic, Autonomous* (Duin & Pedersen, 2021), we emphasize the importance of cultivating the ability to write and work alongside augmentation technologies; this includes working with such non-human agents, understanding the impact of algorithms and AI on work and writing, accommodating the unique relationships with autonomous agents, and planning for ongoing disruption. Again, the purpose of this book is to cultivate an even deeper understanding of human augmentation technology and AI, build technical and professional communication capacity to articulate its benefits and risks, and provide direction for future practice and collaboration.

As Figure 1.3 illustrates, technologies influence cognitive, sensory, physical, and emotional states for the purpose of enhancement, efficiency, and automation, the level of which is increasingly impacted by fluctuating value systems, rhetorical context, and corporate hype. Augmentation technologies function within embodied carryable, wearable, implantable, ingestible, embeddable, robotical, and ambient/spatial platforms. Spurred by artificial intelligence, the resulting autonomous systems and non-human agents will increasingly become part of the fabric of our digital lives, the scope of which is detailed in Chapter 2. Given the proliferation of AI and autonomous systems, we contend that TPCs must prepare for promoting human-autonomous teaming surrounding present and emerging augmentation technologies (see Chapter 7).

Figure 1.3. Technical communication future practice and collaboration

## *Rationale for Technical and Professional Communication Engagement with Augmentation Technologies*

In 2009, Carolyn Rude's seminal work to map research questions in technical communication provided scholars with solid direction for positioning their work. Rude posited this central question for the field of technical and professional communication: "How do texts (print, digital, multimedia; visual, verbal) and related communication practices mediate knowledge, values, and action in a variety of social and professional contexts?" (p. 176). She positioned research within four areas of related questions surrounding disciplinarity ("How shall we know ourselves?"), pedagogy ("What should be the content of our courses and curriculum?"), practice ("How should texts be constructed to work effectively and ethically?"), and social change ("How do texts function as agents of knowledge making, action, and change?").

Amid the current era of superhuman innovation, we propose this central question: How do augmentation technologies (cognitive, sensory, physical, emotional) and related communication practices mediate knowledge, values, and action in professional and personal contexts? Given the proliferation of augmentation technologies, we must articulate expanded disciplinarity as we find ourselves collaborating with non-human agents; we must develop and deploy pedagogy that provides opportunities for such human-autonomous teaming; we must envision texts as being constructed in collaboration with non-human agents, providing TPCs with knowledge to deploy such work effectively, efficiently, and ethically; and we must articulate how AI and augmentation technologies function as agents of knowledge making, action, and change.

In 2009, Andrew Mara and Byron Hawk's special *Technical Communication Quarterly (TCQ)* issue on posthuman rhetorics in technical communication chronicled the philosophical, scientific, sociological, technological, literary, and cultural discourses surrounding posthumanism as a means to bring posthuman perspectives to bear on critical problems in technical communication: workplace identities, organizational situatedness, human–computer interaction, workplace texts and technologies, pedagogical practices, transitioning past/present/future contexts, and organizational change (p. 1). They wrote,

> As organizations become more complex, technologies more pervasive, and rhetorical intent more diverse, it is no longer tenable to divide the world into human choice and technological or environmental determinism. Professional and technical communication is a field that is perfectly situated to address these concerns. Because it is already predisposed to see the writer in larger organizational contexts, the moment is right to explore technical communication's connections to posthumanism, which works to understand and map these complex rhetorical situations in their broader contexts. (p. 3).

More recently, Kristen Moore and Daniel Richard's (2018) collection, *Posthuman Praxis in Technical Communication*, articulated how posthumanism and praxis together offer direction as we grant increased agency to autonomous, nonhuman agents. Carl Herndl's foreword sums it up best: "The shift to a posthuman praxis in technical communication contributes to making what Haraway calls a more lively and livable world. More lively because it is populated by agentive things which escape the modern binary between the human and the nonhuman. More livable because accounting for the actions and possibilities of these newly enfranchised things allows us to better manage the interrelations among this reassembled community" (xiv).

Regarding the future of technical and professional communication, researchers and practitioners also often list emerging trends and associated technologies as a means toward a "more lively and livable world." In his introduction to a special 2005 *Technical Communication* issue on the future of technical communication, Michael Albers argued that "the future of the field will be technology laden. Technology permeates everything a practicing technical communicator does. How we react to changes in that technology on both the individual and organizational level will have a dramatic impact on the development of the profession" (p. 271). Chris Lam (2021), in his recent study of Society of Technical Communication members, networks, trends, and themes from 2016 to 2019 through examination of 75,333 Tweets using the hashtag #TechComm, states that "Technical communication is a field that continually evolves because of its connection to technology—as technology evolves, so too does communication with and about technology. . . . [Its] competencies, skillsets, and areas of expertise, continues to evolve as related fields and sub-disciplines emerge like content strategy and user experience (UX)" (p. 6). Lam articulates dominant themes as including the Adobe TC suite and other authoring tools, the Society of Technical Communication (STC) summit, academic tech comm, professional development, DITA and lightweight DITA, job opportunities in tech comm, and software documentation. Not surprisingly, Lam identifies only limited overlap between academic and practitioner communities, with the word "rhetoric" "tweeted almost exclusively by academics" (p. 17). This aligns with related studies of academic and practitioner journals and magazines (Boettger & Friess, 2016; Andersen & Hackos, 2018). The strongest focus on technology appears within "the practitioner orbit," and Lam identifies XML and DITA instruction as being foundational for the field (note Duin and Tham's 2018 development of such curricula).

Lam also identifies Scott Abel, known as The Content Wrangler, as the practitioner with the largest lively audience: 20,600 followers. In 2019, Abel hosted a four-week Society for Technical Communication Roundtable discussion on the future of technical communication in which participants examined trends, challenges, methods, standards, tools, and technologies in use across technical communication departments that spanned over 600 professionals working in 16 countries. Respondents were "super excited" about the future impact of advanced technologies on the way they live and work, being most excited about chatbots, with 50% of teams planning to launch a bot by 2020, along with voice assistants and AI. They simultaneously anticipated and feared augmented reality (AR), with one director of technical information for a medical device manufacturer stating, "Augmented reality will radically transform how technical and scientific information is communicated, bringing technical documentation content directly to—and integrating it with—the point-of-use."

In 2020, Abel invited Rob Gillespie, information architect from the UK, to share his thoughts in a webinar about the future of technical communication (Zoomin, 2020). Gillespie speaks of the unique role of the technical communicator as a "value generating entity" and the strategic role as technical communication becomes more integrated with and integral to the business. He emphasizes the importance of working closely and concisely with and for machines; being careful about what to input into the machine; and expanding the use of tools to "make us better." He stresses that the "old tools and cubbyholes" are gone and that technical communicators should plan for working with robots and AI. Gillespie promotes three steps for preparing for the future: 1) to be proactive about collaboration, making sure that processes are well defined and efficient; 2) to embrace robotic automation and AI, stating that "they are our friends, [we] must work with them and understand how they work"; and 3) to automate everything that can be automated, emphasizing that technical communicators need to collaborate to move "laborious tasks" to machines.

As a result, while increasing numbers of programs are working to integrate XML and DITA instruction, we see it as imperative for technical communication researchers, instructors, and practitioners to understand how augmentation technologies are modeled to be integrated within our personal and professional lives and proposed to enhance and transform our thinking, sensing, and feeling. It is critical to address the evolving ethical and rhetorical dimensions and dilemmas surrounding them. Therefore, we construct this book as a guide for preparing technical communication practitioners and instructors to cultivate understanding of augmentation technologies, build digital and AI literacy, and design ethical futures. Moreover, we work to provide direction for reframing understanding of audience, usability, and both academic and practitioner roles in technical and professional communication amid the rapid emergence of augmentation technologies.

However, one of the greatest challenges facing technical and professional communication scholars and instructors continues to be a reticence to prepare for such advance of major technological transformations. Citing this urgent need, in *Writing Futures: Collaborative, Algorithmic, Autonomous* (Duin & Pedersen, 2021) we introduced a *Writing Futures* Framework for scholars and instructors to investigate and plan for social, digital literacy, and civic implications of collaborative, algorithmic, and autonomous writing futures. Our goal was to provide readers with opportunities to begin to understand and write alongside non-human agents, examine the impact of algorithms and AI on writing, and accommodate the unique relationships emerging with autonomous agents. *Augmentation Technologies and Artificial Intelligence in Technical Communication* provides greater depth for understanding augmentation technologies and, amid enhancement of professional capabilities, a strong focus on building technical and professional communication ability, and strategic knowledge to articulate its benefits, risks, and relevance.

We understand that readers across our broad technical and professional communication field may well not hold a unified view of professionalization for work with augmentation technologies and artificial intelligence. Saul Carliner (2012) explored such tension, proposing a spectrum from *formal professionalism* "rooted in a worldview that values expertise" to *quasi professionalization* "in which individuals participate in the activities of the occupational infrastructure but without the expectation of exclusive rights to perform the work" to *contraprofessionalization*, in which professionals offer services "outside of parts of the entire infrastructure, sometimes circumventing it completely" (p. 49). At this current time of technological transformation, throughout this book we offer direction for building professionalization, and we agree with Carliner in that such current lack of consensus on the TPC infrastructure for understanding, design, and adoption of augmentation technologies and AI, may well result in "a competitive environment for certain types of high-value assignments" (p. 49) as these technologies emerge. The last decade indeed has further accentuated the need for TPCs to evolve professional understanding of audience, user experience, content management, collaboration, and design.

Moreover, this book involves multiple contours of posthumanism as we increasingly participate in complex assemblages with non-human agents through distributed services and digital platforms. George Hayhoe and Pam Estes Brewer (2021), in *A Research Primer for Technical Communication*, discuss the various goals that direct research, citing Thomas Reeves' (1998) six categories of research: theoretical, empirical, interpretivist, postmodern, developmental, and evaluative. Of these we focus most on postmodern research, "examining the assumptions that underlie applications of technology or technical communication with the ultimate goal of revealing hidden agendas and empowering disenfranchised groups" (Hayhoe & Brewer, p. 7).

Table 1.1 provides an initial summary of how core functions of technical communication— audience, usability, content development, collaboration, and design—transform alongside augmentation technologies. These core functions evolve as AI and use of non-human agents become increasingly integrated throughout TPC work. With such integration, we need to increase focus on the audience as immersed (Tham, 2018) and now augmented through increased awareness as a result of augmentation technology use. User experience and usability testing evolve to include augmentation technology support and increased use of non-human agents as team members. Content management systems and associated standards (e.g., DITA) will include increased use of machine learning and explainable AI; and collaboration will increasingly depend on human-autonomy teaming.

| Core function | Current | Future |
|---|---|---|
| Audience addressed, invoked, involved | Audience immersed | Audience [agency] augmented |
| User/human experience (UX) | Augmentation technology as UX team support, interdependent team member | Non-human agent as independent team member, digital employee |
| Content development | Topic-based content management systems | Machine learning and explainable AI |
| Collaboration | Participatory design <br><br> Agile development | Human-autonomy teaming |
| Dialogic design | Social justice design ethic <br><br> Surveillance culture | AI ethics <br><br> Algorithmic impact assessment |

Table 1.1. Current and future core functions of technical and professional communication.

Throughout this evolutionary change, TPCs will continue to be tasked with understanding and documenting augmentation technologies to make them understandable and usable by professionals and the public. Too often TPCs are provided with a combination of marketing information and/or technical specifications from subject matter experts, with neither set of documents correctly or adequately representing what users need.

Related disciplines are studying this evolutionary change. Shyam Sundar and Eun-Ju Lee (2022) edited a recent issue of *Human Communication Research* in which contributing authors explored the role of AI in communication along conceptual dimensions in human–computer interaction (HCI) and computer-mediated communication (CMC) research, describing how AI can fulfill analogous roles of communicator or mediator. They ask, "How well can AI replace a human in serving as a communicator?" Here we explore augmentation technologies and AI, with focus on human–AI collaboration. Throughout this book, we describe how technical and professional communication (TPC) roles will change as a result of increased use of augmentation technologies and AI. We agree with Sundar and Lee in that AI technology "has profoundly affected virtually all areas of our lives over the past decade" (p. 379) In their case, the traditional distinctions between HCI and CMC now must be modified, given AI involvement. In our case, traditional approaches to audience, purpose, and intended effects now must be modified. Their special issue was conceived "to address the questions of how such wide- spread integration of AI in communication might alter the very essence of human communication and force communication scholars to revisit widely accepted assumptions and knowledge about how we humans communicate, with what consequences" (p. 381). In this book, we provide rationale and direction for positioning TPCs to understand augmentation technologies and AI as a means to design ethical futures across this work.

Moreover, the Fabric of Digital Life, or "Fabric" (https://fabricofdigitallife.com), detailed in Chapter 2, provides a means to access and understand the metadata behind an emerging technology, resulting in deeper knowledge of modes of technology invention over time and how technologies move from the seeming fringe to mainstream use. We advocate for greater understanding of metadata as a means for TPCs to better understand augmentation technologies, resulting in more informed persona development and information design. Examining metadata allows us to uncover sociotechnical tradeoffs in the technologies (Iliadis & Pedersen, 2018). The burgeoning augmentation technologies market masks complex sociotechnical tradeoffs that TPCs must understand as part of their work to develop usable and ethical content. For example, a business will frame an augmentation technology as a business solution without conveying critical information about the concessions that users must make in terms of giving up their data in exchange for access and use. TPCs can examine Fabric's rich metadata fields to search for terms and technologies as a means to better understand augmentation technologies and the potential tradeoffs related to usability.

Rhetorical scholar K. J. Rawson (2018) explored how such curated archival description results in greater bureaucratic and epistemological understanding, providing an extended consideration of the *description* archival process. Rawson quotes MacNeil (2009, p. 90) stating that archival description "involves telling a story about records, which both 'changes the meaning of the records' and 'determine[s] how they will be used and remade in the future.' Such transformations evidence the rhetorical power of archival description. . . . It is created to suit particular audiences and needs; and it can have tremendous influence over the reception and use of the materials it describes" (pp. 328–329). Thus, the archival descriptions in Fabric, i.e. the collection descriptions and the metadata for each digital artifact, provide tremendous understanding of the reception and use of augmentation technologies. TPCs are well positioned to understand metadata as a means to unveil and address the intricacies of augmentation technologies. In Chapter 2 we illustrate how we use Fabric of Digital Life metadata to help define augmentation technologies, trace them over time, and provide a means to prepare for the present and the future.

## *Augmentation Technologies—Designing Ethical Futures Framework*

The core of our human experience, identity, and professional and personal practices is affected by the momentum of augmentation technologies and AI that enhance human capabilities or productivity by adding to the body (or ambient environment around the body) cognitive, physical, sensory, and/or emotional enhancements. Emerging wearable devices give people with sight loss the ability to see their environment; earbuds translate conversations in real time; nanobots deployed into the body deliver drugs to target and attack disease; and neural implants transmit brain activity to allow humans to control machines using only thoughts.

This book brings theoretical, empirical, pedagogical, critical, and ethical attention to the development of these sophisticated, emergent, and embodied augmentation technologies slated to promote enhanced futures—i.e., to improve lives, literacy, cultures, arts, economies, and social and professional contexts. To date, the study of the emergence of augmentation technologies largely has neglected to adopt such an ethical, human-centric approach, leaving citizens, instructors, practitioners, and governments to deal with the consequences. Moreover, while technical and professional communication has begun to come to terms with socio-ethical problems and inequities (Walton, Moore, & Jones, 2019), augmentation technology and AI design, adoption, and adaptation have not followed suit, nor have they properly dealt with grand challenges including algorithmic bias, lack of diversity in development teams, and misuse of augmentation technologies and AI in the field.

Figure 1.4 provides an overview of the major sections and associated themes in this book. As shown in the center of this visual, the purpose of this work is to reframe professional practice and pedagogy to promote digital and AI literacy surrounding the ethical design, adoption, and adaptation of augmentation technologies. The three larger shaded boxes indicate the major sections in the book; the lighter shaded boxes indicate key themes presented in each section:

- Understand (rhetorics of) augmentation technologies

  - Introduces the Ethical Futures Framework, defines and articulates the dimensions of augmentation technologies, focuses on the development of non-human agents in industry as a critical factor in the rise of augmentation technologies, and provides critical attention to hyped assumptions and recent claims concerning the Metaverse and AR.

- Build literacies

  - Provides detailed definition of digital and AI literacy amid the emergence of augmentation technologies, with focus on examining, exploring, and participating in the development of digital and AI literacy skills needed to address algorithmic mining and bias, racial discrimination, digital divides, unethical AI practices, misinformation, and other socio-ethical harms to humans.

- Design ethical futures

  - Provides direction for cultivating digital literacy and AI literacy skills needed to assess integration and participation with augmentation technologies, chronicles the rapid increase in autonomous agents and digital employees, positions TPCs for intervening throughout the design, adoption, and adaptation of augmentation technologies, and shares strategic and tactical approaches for designing ethical augmentation technology and AI futures.

Figure 1.4. Ethical Futures Framework

Only by understanding and embedding ethical principles throughout augmentation technology design and AI can we foster human-centered, humane futures. As we detailed in *Writing Futures*, the Berkman Klein Center at Harvard University started the Principled Artificial Intelligence Project as a means to map ethical and human rights-based approaches to AI (Fjeld & Nagy, 2020). The extensive data visualization synthesizes 36 principles documents, focusing on eight themes: Promotion of Human Values, Professional Responsibility, Human Control of Technology, Fairness and Non-Discrimination, Transparency and Explainability, Safety and Security, Accountability, and Privacy. This project found that stakeholders do not necessarily agree on how to define and apply responsible AI principles. They write, "Despite the proliferation of these 'AI principles,' there has been little scholarly focus on understanding these efforts either individually or as contextualized within an expanding universe of principles with discernible trends." We draw on these key themes throughout this book, contextualizing them for technical and professional communication.

This book also draws on government, private, and civil sectors, Inter-Governmental Organizations, and academic groups such as the IEEE Ethically Aligned Design initiative, analyzing and illustrating that stakeholders do not always agree; e.g., some entities emphasize human control of technology and others ignore it. Therefore, we find the Principled Artificial Intelligence visualization to be extremely useful in research, teaching, and outreach, as it illustrates how AI values can be seen as heterogeneous, dynamically changing, and always contextualized. Figure 1.5 from the top of this visualization includes the key themes and reports from civil society and government organizations, themes that we discuss in detail in Chapter 5.

Figure 1.5. A section from the Principled Artificial Intelligence visualization, Berkman Klein Center at Harvard University. Creative Commons Attribution 3.0 Unported license. https://creativecommons.org/licenses/by/3.0/

Indeed, researchers in AI ethics, robot ethics, and philosophy are actively delineating value-based frameworks for embedding synthetic emotions in augmentation technologies. Throughout this book, we also invite you to use the key IEEE resource *Ethically Aligned Design: A Vision for Prioritizing Human Wellbeing with Artificial Intelligence and Autonomous Systems, First Edition*. As the report states, its purpose is "to establish frameworks to guide and inform dialogue and debate around the non-technical implications of these technologies, in particular related to ethical aspects. We understand 'ethical' to go beyond moral constructs and include social fairness, environmental sustainability, and our desire for self-determination" (The IEEE Global Initiative on Ethics of Autonomous and Intelligent Systems, 2019, p. 3).

## Sections and Chapters

The sections of this book align with the three sections of the Ethical Futures Framework. Section 1 concentrates primarily on understanding the rhetorics of augmentation technologies. In this chapter we have provided an initial definition of augmentation technologies along with prominent scholarly directions that inform its development. We have included rationale for positioning technical and professional communication researchers, instructors, and practitioners to understand and articulate future changes to core technical communication functions of audience, user experience, content development, collaboration, and design as impacted by augmentation technologies and AI.

Chapter 2 identifies a changing set of value systems that constitute augmentation technologies as social phenomena, namely, beliefs surrounding enhancement, automation, and efficiency. It articulates and examines cognitive, sensory, emotional, and physical enhancements as a range of subcategories of augmentation technologies and AI. It illustrates how the Fabric of Digital Life (https://fabricofdigitallife.com) archive's rich metadata fields provide a means to examine augmentation technologies and the complex sociotechnical tradeoffs that technical and professional communicators must navigate as they work to develop usable content and direction.

Chapter 3 provides critical attention to the hyped assumption that sophisticated, emergent, and embodied augmentation technologies will improve lives, literacy, cultures, arts, economies, and social contexts. It begins with discussion of the problem of ambiguity with AI terminology, providing a description of the WIPO Categorization of AI Technologies Scheme to assist with it. This chapter then draws on media and communication studies to explore concepts such as agents, agency, power, and agentive relationships between humans and robots. The chapter focuses on the development of non-human agents in industry as a critical factor in the rise of augmentation technologies. It looks at how marketing communication enculturates future users to adopt and adapt to the technology. Scholars are charting the significant ways that people are drawn further into commercial digital landscapes, such as the Metaverse concept, in post-Internet society. It concludes by examining recent claims concerning Metaverse and AR through the use of an electronic soft contact lens platform.

Section 2 concentrates primarily on pedagogy and guiding principles for building digital and AI literacy. Chapter 4 defines digital and AI literacy for current and future work with augmentation technologies. The exponential increase in AI techniques, functional applications, and use across application fields demands critical attention to digital literacy, data literacy, AI literacy, AI explainability, and trustworthy AI. This chapter uses Long and Magerko's (2020) definition of AI literacy and conceptual framework for determining AI competencies and design considerations as a means for TPC scholars, instructors, students, and practitioners to examine and develop digital and AI literacies. It focuses on explainable AI (XAI) as a means to help humans better understand how AI works and makes particular decisions. It then discusses the Assessment List for Trustworthy AI (ALTAI) found at the European AI Alliance portal (Welcome to the ALTAI portal!, 2020), a set of strategic questions for use as an initial approach to evaluating trustworthy AI for the purpose of minimizing risks while maximizing the benefits of AI for human users. The chapter concludes with discussion of how technical and professional communication (TPC) roles are evolving as a result of augmentation technologies and AI.

Chapter 5 addresses potential socio-ethical consequences of augmentation technologies and AI. It provides clarity on biometric data and its relevance to augmentation technologies in combination with artificial intelligence. *Principled Artificial Intelligence: Mapping Consensus in Ethical and Rights-Based Approaches to Principles for AI* (Fjeld et al., 2020) provides the ethical focus. The chapter uses work published by the AI Now Institute, which monitors and reports on human rights issues through several published reports. These include focus on algorithmic mining and bias, racial and gender discrimination, ableism, digital divides, unethical AI practices, misinformation, and other socio-ethical harms to humans. We continue to build rationale for understanding digital and AI literacy as a means to avoid socio-ethical harms to humans that occur when digital and AI literacy is absent.

Section 3 concentrates primarily on practitioner direction, beginning in Chapter 6 with discussion of pedagogical direction for cultivating digital literacy and AI literacy skills needed to assess integration and participation with augmentation technologies. We draw on multiple studies on building digital literacy (Burnham & Tham, 2021; Davis et al., 2021; Duin et al., 2021; Tham et al., 2021) to provide direction for cultivating digital and AI literacy through examining and curating augmentation technologies. We describe student and instructor perception of digital and AI literacy through studies of learner engagement with the Fabric of Digital Life archive/artifacts along with theoretical and pragmatic model development for instructor-scholars to design and integrate digital and AI literacy development that caters to student learning needs as well as their professional futures and workforce preparedness. The chapter provides suggestions for integrating study of augmentation technologies in technical communication across undergraduate and graduate curricula, concluding with discussion of learner capabilities for a world with augmentation technologies and AI.

Chapter 7 begins with discussion of guidelines for professional practice surrounding human–AI interaction that includes a TPC guide to human–AI interaction based on key studies in human-centered computing, science and engineering, and technical communication. Given the rapid increase in autonomous agents and human–AI interaction, the chapter's main focus is on seeing tools as autonomous agents or digital employees. After discussion of chatbots and conversational design, the chapter chronicles the development of digital employees across six corporations. Based on AI functional technologies, these digital employees are proposed to enhance meaningful, empathetic connections to the digital world, resulting in potentially greater symbiotic relationships with users. TPCs must be positioned to intervene throughout the design, adoption, and adaptation of augmentation technologies. The audience is now an "augmented" audience; autonomous agents now function as independent team members; and content is produced through machine learning.

Chapter 8 focuses on designing ethical futures by way of strategic and tactical approaches to governance, regulation, and standardization of augmentation technologies and AI. Influenced by philosopher Michel de Certeau's (1984) distinction between strategies and tactics, the chapter concentrates on TPC capabilities to enact changes in their field toward ethical ends. The chapter begins by noting exemplary organizations working to promote ethical design of augmentation technologies, whose collective policy guidelines work to address risk. It then focuses on risk communication and awareness of ethical algorithmic impact assessment tools and processes to help guide design of and collaboration with augmentation technologies, including digital employees. Strategic approaches include work underway in the European Union and proposed in the US. Tactical approaches discussed include coalition and standards building, including Metaverse professional standards bodies currently forming to help build foundations for open standards among corporate stakeholders, and a journey mindset. The chapter concludes with an invitation to collaborate on research underway as part of the Digital Life Institute at https://www.digitallife.org/.

To conclude, professionals across multiple industries, including at IBM (IBM, 2022), Microsoft (Spencer, 2019), and Samsung (Samsung Advanced Institute of Technology, n.d.), welcome multi-disciplinary direction for navigating the evolving augmented technology landscape. For example, Samsung is proposing an exoskeleton technology concept for fitness and entertainment to be used in the home. Samsung's Innovation Campus (SIC) offers AI development education and teaching materials to youth around the world, and the annual CES trade show, "the world's most influential technology event," promotes and celebrates augmentation technologies far in advance of their actual deployment. While futuristic model cars and future homes are often on display, so too are human augmentation technologies. The January 2020 CES featured a session called "The Next Era: Superhuman Innovation" asking "What if we could tap into superhuman powers to be better at school, excel at sports, succeed in business and ultimately live a longer and fuller life? Who wouldn't want that competitive edge?" (Duffey, 2019). And the January 2022 CES featured sessions on biowearables of the future, smart contact lenses, and access to the Metaverse via Vuzix Shield AR glasses.

Moreover, corporations seek direction as they are tasked with requirements to develop algorithmic impact assessment tools and processes to ensure transparency and accountability for decision-making systems (Reisman et al., 2018). The US Algorithmic Accountability Act of 2022, as summarized at https://www.wyden.senate.gov, "requires companies to assess the impacts of the automated systems they use and sell, creates new transparency about when and how automated systems are used, and empowers consumers to make informed choices about the automation of critical decisions." Working in tandem with usability experts, with this book's direction, technical and professional communication scholars and practitioners will be positioned to meet increased expectations for the ethical design, adoption, and adaptation of augmentation technologies and AI.